      \documentstyle[12pt,graphicx]{article}                
\begin{document}
\baselineskip 18pt
\def\today{\ifcase\month\or
 January\or February\or March\or April\or May\or June\or
 July\or August\or September\or October\or November\or December\fi
 \space\number\day, \number\year}
\def\thebibliography#1{\section*{References\markboth
 {References}{References}}\list
 {[\arabic{enumi}]}{\settowidth\labelwidth{[#1]}
 \leftmargin\labelwidth
 \advance\leftmargin\labelsep
 \usecounter{enumi}}
 \def\newblock{\hskip .11em plus .33em minus .07em}
 \sloppy
 \sfcode`\.=1000\relax}
\let\endthebibliography=\endlist
\def\lsim{\ ^<\llap{$_\sim$}\ }
\def\gsim{\ ^>\llap{$_\sim$}\ }
\def\r2{\sqrt 2}
\def\beq{\begin{equation}}
\def\eeq{\end{equation}}
\def\beqn{\begin{eqnarray}}
\def\eeqn{\end{eqnarray}}
\def\rmuu{\gamma^{\mu}}
\def\rmud{\gamma_{\mu}}
\def\PL{{1-\gamma_5\over 2}}
\def\PR{{1+\gamma_5\over 2}}
\def\sinW2{\sin^2\theta_W}
\def\AEM{\alpha_{EM}}
\def\mul{M_{\tilde{u} L}^2}
\def\mur{M_{\tilde{u} R}^2}
\def\mdl{M_{\tilde{d} L}^2}
\def\mdr{M_{\tilde{d} R}^2}
\def\mz2{M_{z}^2}
\def\c2b{\cos 2\beta}
\def\au{A_u}         
\def\ad{A_d}
\def\cob{\cot \beta}
\def\v#1{v_#1}
\def\tb{\tan\beta}
\def\epem{$e^+e^-$}
\def\KK{$K^0$-$\bar{K^0}$}
\def\wi{\omega_i}
\def\xj{\chi_j}
\def\Wmu{W_\mu}
\def\Wnu{W_\nu}
\def\m#1{{\tilde m}_#1}
\def\mH{m_H}
\def\mw#1{{\tilde m}_{\omega #1}}
\def\mx#1{{\tilde m}_{\chi^{0}_#1}}
\def\mc#1{{\tilde m}_{\chi^{+}_#1}}
\def\mwi{{\tilde m}_{\omega i}}
\def\mxi{{\tilde m}_{\chi^{0}_i}}
\def\mci{{\tilde m}_{\chi^{+}_i}}
\def\mz{M_z}
\def\sw{\sin\theta_W}
\def\cw{\cos\theta_W}
\def\cb{\cos\beta}
\def\sb{\sin\beta}
\def\rwi{r_{\omega i}}
\def\rxj{r_{\chi j}}
\def\rfp{r_f'}
\def\Kik{K_{ik}}
\def\Fq2{F_{2}(q^2)}
\def\f{\({\cal F}\)}
\def\d1{{\f(\tilde c;\tilde s;\tilde W)+ \f(\tilde c;\tilde \mu;\tilde W)}}
\def\tw{\tan\theta_W}
\def\sec2w{sec^2\theta_W}

\begin{titlepage}
\begin{center}
{\large {\bf  PHYSICS FROM EXTRA DIMENSIONS }}\\
\vskip 0.5 true cm
\vspace{2cm}
\renewcommand{\thefootnote}
{\fnsymbol{footnote}} Pran Nath$^{a,b,c}$  
\vskip 0.5 true cm
\end{center}

\noindent 
{a. Department of Physics, Northeastern University,
Boston, MA 02115-5000, USA\footnote{: Permanent address}} \\
{b. Physikalisches Institut, Universitat Bonn, 
Nussallee 12, D-53115 Bonn, Germany}\\
{c. Max-Planck-Institute fuer Kernphysik, Saupfercheckweg 1,
D-69117 Heidelberg, Germany}\\

\vskip 1.0 true cm
\centerline{\bf Abstract}
\medskip
A brief review of the recent developments 
 in the physics from extra dimensions is given with a focus on the effects
of Kaluza-Klein excitations in the Standard Model sector. It is shown that 
the current accurate data on the Fermi constant and on other 
electro-weak parameters puts a lower bound 
on the scale of extra dimensions of $\sim$ 3 TeV, and thus  
the observation of such dimensions lies beyond the reach of
accelerators in the near future. The correction to the anomalous magnetic 
moment of the muon from extra dimensions is discussed and one finds that with 
the current limit on the scale of extra dimensions from the Fermi constant, 
the correction to $g_{\mu}-2$ does not 
compete with the potentially large contributions from the supersymmetric  
electro-weak correction. The possibility of generating Kaluza-Klein
excitations associated with large radius compactifications at the
LHC is discussed. It is shown that if such excitations are indeed
produced their resonance structure will encode information on the
number of compactified dimensions as well as on the 
nature of the specific orbifold compactification. 
 A brief discussion of difficulties such as rapid proton
decay that one encounters in theories with large radius compactifications
is  given.
\end{titlepage}

\section{Introduction}
The physics of extra dimensions has a long and interesting 
history\cite{kaluza,appel,a} beginning with the work of Kaluza and Klein in the 
nineteen twenties\cite{kaluza,appel}.
Recent interest in Kaluza-Klein theories arises because
such theories may arise in TeV scale strings\cite{horava,tye,kaku}. 
Activity 
in these models in taking place along three directions:
(a)Effects of large extra dimensions in the Standard Model 
sector\cite{a,d,gf}, (b) effects in the gravitational 
sector\cite{a1}, and (c)
non-factorizable geometries\cite{gogber,rs}. 
The focus of this talk is on the constraints on
extra dimensions from precision electro-weak data.
Specifically we will discuss in detail the constraints arising
from two of the most precisely determined quantities in 
all of particle physics, i.e., the Fermi constant and the
anomalous magnetic moment of the muon. We will also discuss the
possible signatures for extra dimensions in pp collisions at the
Large Hadron Collider (LHC). Although the focus of this paper 
is on extra dimensions in the context of the Standard Model sector,
we will make a brief detour to assess also the status of 
work on extra dimensions in low scale quantum gravity. 
Finally, We will discuss some of the 
difficulties that surface in theories with
extra dimensions. The outline of this write up is as follows:
In Sec.2 we give a brief discussion of the conventional string
phenomenology based on heterotic strings. In Sec.3 we discuss
the more recent developments which lead to the possibility of
a string scale in the TeV region. In Sec.4 we discuss  
compactifications of extra dimensions which lead to 
Kaluza-Klein excitations in the Standard Model sector. In Sec.5 
we discuss the contributions from the Kaluza-Klein excitations on
the Fermi constant and give an analysis of the constraints that 
the accurate determination of the Fermi constant places on the
compactification scale $M_R$. In Sec.6 an analysis of the
effects of Kaluza-Klein excitations on the anomalous magnetic
moment of the muon is given. In Sec.7 we discuss the probe of 
extra dimensions at colliders. A brief discussion of low scale
quantum gravity is given in Sec.8. In Sec.9 we discuss 
 the difficulties encountered in models with large radius
compactifications. Conclusions are given in Sec.10.

\section{Conventional String Phenomenology}
Extra space time dimensions are an integral part of 
string theory. However, in conventional string phenomenology 
compactifications of extra dimensions occurs at a high scale
close to the 4 dimensional Planck scale $M_{Pl}=1.2\times 10^{19}$
GeV. The effects of extra dimensions in this case are 
  at the level of threshold effects of heavy states with
masses typically order $10^{17}$ GeV.
Such models, based on compactifications on Calabi-Yau 
manifolds, orbifolds, and free fermionic constructions possess
many desirable features\cite{stringmodels}. 
Thus they contain the Standard Model 
gauge group, N=1 supersymmetry and can accomodate the spectrum
of the Minimal Supersymmetric Standard Model (MSSM). In these
models there is an automatic unification of the gauge coupling
constants at the scale $M_{str}$. Unfortunately, compatibility 
with the LEP data is not guaranteed. In fact, one has to invoke 
the presence of additional phenomena  in the form of either
 an additional set of states over and above the MSSM 
spectrum with intermediate scale masses or large
threshold corrections to get agreement with the LEP data. 
Further, in some models there is the
problem of extra light Higgs doublets and the problem of 
proton stability in some others. Nonetheless it is quite 
remarkable that several string models come close to becoming realistic.
Since supersymmetric grand unification is very successful in 
accomodating the unification of gauge couplings  given
by the LEP data, recent efforts have focussed on deducing 
grand unification from higher level Kac Moody levels\cite{kac2}. 
However, some phenomenological problems still remain to be 
resolved in these constructions.

\section {Recent String Model Building}
Recent developments in string model building has proceeded along
two main directions: (i) M theory compactification, and (ii) Type I 
[Type IIB] string compactifications. The generic features of 
such models is that there is no longer a rigid relation 
between $M_{str}$ and $M_{Planck}$\cite{horava}. 
In fact in the context of
Type I[Type IIB] compactifications, the scale $M_{str}$ can 
be as low as a TeV. 
We shall discuss models where the string 
scale and the compactification scale are indeed quite low, i.e.,
in the TeV range. 
In addition to $M_{srt}$ being low the fundamental
scale of gravity in higher dimensions may be low\cite{a1}. 
This is possible because the observed Planck scale $M_{Pl}$ 
in four dimensions and the fundamental gravity scale in
higher dimensions are related by extra dimensions\cite{a1}.
We shall discuss the implications of a low scale quantum gravity
in further detail in Sec.8.

\section {TEV Scale Strings and Kaluza-Klein Modes} 
In this section we  discuss the effects of Kaluza-Klein 
  modes in the Standard Model sector in  models with large 
radius compactification. 
The simplest phenomenologically viable example of a higher dimensional
theory is the case with one extra dimension which we assume is 
compactified on $S^1/Z_2$ with a 
compactification radius of $R=M_R^{-1}$. 
After compactification  the resulting spectrum contains massless 
modes with N=1 supersymmetry in 4D, which precisely form the spectrum 
of MSSM in 4D. The  massive Kaluza-Klein modes form N=2 multiplets in 4D
with masses $(m_i^2+n^2 M_R^2), n=1,2,3,.., \infty$ where
 $m_i^2$ is the electro-weak mass and $n^2M_R^2$ terms is the
compactification mass. For simplicity we consider a direct 5 dimensional 
extension of the MSSM with the matter fields (quarks, leptons and
Higgs) confined to the orbifold points which constitutes the 
4 dimensional wall of the physical space time, while the
$SU(3)_C\times SU(2)_L\times U(1)$ gauge bosons propagate in the
bulk. In this model after compactification of the fifth dimension
we will have only zero modes for the matter fields while the gauge 
bosons will contain both the zero modes as well as the Kaluza-Klein
 modes. In the effective theory in 4D a rescaling of the 
five dimensional gauge coupling constant is necessary 
 so that $g^{(5)}_i/{\sqrt{\pi R}}=g_i$.
After rescaling the low energy effective lagrangian in 
4 dimensions is of the form 

\begin{equation}
{\it L_{int}}=g_ij^{\mu}(A_{\mu i}+\sqrt 2\sum_{n=1}^{\infty} A_{\mu i}^n)
\end{equation}
where $A_{\mu i}$ are the gauge fields and are massless,
$ A_{\mu i}^n$ are their massive Kaluza-Klein excitations,
and $j_{\mu}$ are the matter sources which contain the massless
  quark, leptons and Higgs fields.
It is interesting to note that the coupling of the  vector Kaluza-Klein 
modes to matter is
a factor $\sqrt 2$ larger than the coupling of the zero mode to
matter.  The above compactification scheme can be generalized to
the case with more than one extra  dimension. However, as
the number of extra dimensions becomes larger 
the number of possible compactifications also grows. Thus, for
example, for the case of two extra dimensions on may compactify
on  $Z_2\times Z_2$, $Z_3$ and $Z_6$ orbifolds. In general,
different compactifications will lead to different low energy
effective 4D theories and to different signatures in low energy
 physics. We will return to this topic in Sec.7.

\section{ Kaluza-Klein Effects on the Fermi Constant}
The Fermi constant is very accurately known from the muon lifetime.
 From the complete 2 loop corrections one has\cite{ritbergen}
\begin{equation}
 G_F=1.16639(1)\times 10^{-5} GeV^{-2}
\end{equation}
A comparison of the Standard Model value with the experimental
value of Eq.(2) shows an excellent agreement between theory
and experiment. However, the  
error in the theoretical determination of $G_F^{SM}$ 
is much larger, by a factor of
around two orders of magnitude, than the error in the 
experimental determination given by Eq.(2). 
It is this theoretical error that allows for 
the possibility of a correction from the Kaluza-Klein states.
Specifically, the Kaluza-Klein correction to the Fermi constant 
 must lie in the error corridor of the experimental value
 and the Standard Model prediction, i.e.,  
$\Delta G_F^{KK}/G_F^{SM}=G_F/G_F^{SM}-1$.
 Thus for d extra dimensions  
the effective Fermi constant including the Kaluza-Klein corrections
is given by\cite{gf} 
\begin{equation}
G_F=G_F^{SM} \int_{0}^{\infty}dt e^{-t}  
(\theta_3(\frac{itM_R^2}{M_W^2\pi}))^{d}
\end{equation}
 where $\theta_3(z)$ is the Jacobi function defined by 
$\theta_3(z)=\sum_{k=-\infty}^{\infty} e^{(i\pi k^2z)}$.
For the case of one extra dimension one finds  to leading order in
$M_W/M_R$ the result\cite{gf}
\begin{equation} 
G_F^{eff}\simeq G_F^{SM}(1+\frac{\pi^2}{3}\frac{M_W^2}{M_R^2})
\end{equation}
Thus in the case of one extra dimension 
$\Delta G_F^{KK}/G_F^{SM}$=$\frac{\pi^2}{3}\frac{M_W^2}{M_R^2}$
and a direct determination of $M_R$ is possible from the 
error corridor between  $G_F$ and $G_F^{SM}$. However, the error
corridor which is essentially determined by the error in
$G_F^{SM}$ is very sensitively dependent on the scheme in which
radiative corrections are computed as well as on the process used 
to extract it. We illustrate this with two parametrizations of 
$G_F^{SM}$. First in the on-shell scheme one has 

\begin{equation}
G_F^{(SM)}=\frac{\pi \alpha}{\sqrt 2M_W^2 \sin^2\theta_W(1-\Delta r)}
\end{equation}
where  $\sin^2\theta_W=(1-M_W^2/M_Z^2)$ and $\Delta r$ is the
radiative correction in this scheme. Alternately one may parametrize
 $G_F^{(SM)}$  by 
\begin{equation}
G_F^{(SM)}=\frac{\pi\alpha}{\sqrt 2M_Z^2\hat s^2\hat c^2 
(1-\Delta\hat r)}
\end{equation}
where $\hat\Delta r$ is the radiative correction and 
$\hat s=\sin\theta_W(\bar{MS})$ 
and $\hat c=\cos\theta_W(\bar {MS})$.
Several other determinations of $G_F^{SM}$ exist
as well, e.g., from the leptonic partial decay widths of the
Z boson (see, e.g., Marciano in Ref.\cite{gf}). 
With these parametrizations and the current errors in the 
electro-weak parameters one finds $M_R\geq 3$ TeV 
with a $\pm 1$ TeV fluctuation depending on the parametrization
used.
With the above limit on $M_R$ none of the Kaluza-Klein excitations 
of $\gamma$,  W or Z boson will become visible at the Tevatron.
However, the current limit on $M_R$ still allows for 
the possibility that these excitations may become visible at the LHC. 

\section{Fermionic Moments}

The extra dimensions will also have an effect on the fermionic moments,
and specially on the muon anomalous magnetic moment.
This moment is one of the most accurately
determined quantities in physics. The most recent measurement of
$a_{\mu}=(g_{\mu}-2)/2$ from the Brookhaven experiment E821 combined
with the previous CERN measurement\cite{bailey} 
gives\cite{brown,marciano}
\begin{equation}
a_{\mu}^{exp}=11659235.7(46)\times 10^{-10}
\end{equation}
 The result
of the Standard Model for this quantity computed to  $\alpha^5$ 
QED corrections\cite{kinoshita} and up to $\alpha^2$ hadronic\cite{hocker} 
and two loop
electro-weak corrections gives the value\cite{marciano} 
\begin{equation}
a_{\mu}^{SM}=11659159.7(6.7)\times 10^{-10}
\end{equation}
The two loop electro-weak correction by itself is
$a_{\mu}^{EW}(2~loop)=15.2(0.4)\times 10^{-10}$\cite{czar}. 
It is expected
that in the next round of analysis the BNL experiment will 
reach a sensitivity of $\sim \pm 15 \times 10^{-10}$ and 
eventually it will 
 measure  $a_{\mu}$ to a sensitivity of $\pm 4\times 10^{-10}$.
 $a_{\mu}$ is generally regarded as a sensitive probe 
of new physics and we wish to determine if this is  
also the case for extra dimensions. Thus it is
already known that $a_{\mu}$ is a sensitive probe of 
supersymmetry\cite{yuan} specifically for the case of large $\tan\beta$
since in SUSY $\Delta a_{\mu}\sim \tan\beta$ and for large $\tan\beta$ 
the BNL experiment in fact can favorably compete with the
Tevatron in discovering new physics\cite{yuan}.

 Next we discuss the effects on  $a_{\mu}$ due to corrections from
the Kaluza-Klein excitations of the photon and of the W and 
Z bosons\cite{gmu}. 
In the analysis 
we have to take into account the effects of the Kaluza-Klein W states 
on $G_F$ which we have already discussed. Including these effects
we find that for the case d=1 the correction to $a_{\mu}$  
due to the Kaluza-Klein excitations of $\gamma$, W, and Z
 is given by\cite{gmu} 

\begin{eqnarray}
(\Delta a)_{\mu}^{\gamma-W-ZKK}=\frac{G_F m_{\mu}^2}{\pi^22\sqrt 2}
(-\frac{5}{12}+\frac{4}{3}(sin^2\theta_W-\frac{1}{4})^2)
 (\frac{M_Z^2-M_W^2}{M_R^2})
+\alpha \frac{\pi}{9}\frac{m_{\mu}^2}{M_R^2}
\end{eqnarray}
From Eq.(9) one finds that there is a partial 
 cancellation between the W and Z Kaluza-Klein exchange 
contribution and the photonic Kaluza-Klein exchange contribution. 
The net result is that with $M_R\geq 1$ 
TeV the contribution of the Kaluza-Klein modes to $a_{\mu}$ falls  more than 
 1-2 orders of magnitude below the sensitivity that will be achievable 
in the new $a_{\mu}$ experiment.
Thus extra dimensions even as low as 1 TeV provide no serious 
background for SUSY effects and if a deviation in $a_{\mu}$ 
from the Standard Model prediction is seen at BNL it could not 
be attributed to  effects of extra dimensions.

 The basic reason why there is a very large suppression of the Kaluza-Klein 
contributions to $a_{\mu}$ is because of the redefinition of the 
Fermi constant which absorbs most of the correction in $a_{\mu}$
 from Kaluza-Klein  modes. As discussed in the first paper 
 in Ref.\cite{gf} there is,
however, a variant model where muon decay and  consequently $G_F$
 receive no contribution from the Kaluza-Klein excitations. 
This is a model where the first quark
lepton generation lies in the bulk while the second generation
lies on the 4D wall. In this model while the correction to
$G_F$ from the Kaluza-Klein states are suppressed there is no
such suppression for the Kaluza-Klein correction to 
$a_{\mu}$. The detailed analysis here shows that the new BNL experiment
 will be able to probe extra dimensions for this model as follows\cite{gmu}:
            $M_R\sim 0.65 ~TeV ~(d=2)$,
             $ M_R\sim 1~TeV~(d=3)$, and 
             $M_R\sim 1.4~(d=4)~ TeV$.
  The effects of
 extra dimensions in the context of quantum gravity is
discussed in Ref.\cite{graesser}. With the current estimates on
the scale of low scale quantum gravity (see section 8) 
the quantum gravity effects on $a_{\mu}$ are again expected to be
rather small. Thus aside from the special case of the
variant model discussed above one 
finds that the effect of extra dimensions on $a_{\mu}$ 
will in general be 
too  small to provide any serious
 background to the supersymmetric electro-weak  correction.

\section{Probe of Extra Dimensions at Colliders}
If the compactification radius of extra dimensions is large enough,
the Kaluza-Klein excitations of the $\gamma$, W and Z could be
produced at the Large Hadron Collider [LHC]\cite{sst,abq,nyy}.
In this case one can show that quite remarkably the experimental
data on the Kaluza-Klein excitations encodes information on the nature of
compactification\cite{nyy}. Thus the resonance structure in the production
cross-section associated with Kaluza-Klein excitations will provide 
information on the number of compactified dimensions as well as
on the nature of the specific orbifold compactification.
The most dramatic signals arise from the interference pattern involving 
the exchange of the Standard Model
spin 1 bosons  ($\gamma$ and Z) and their Kaluza-Klein modes.
Additional signals arise from the Kaluza-Klein
 excitations of the W boson and of the gluon. 

\begin{figure}[t]
 \vspace{8.0cm}
\includegraphics{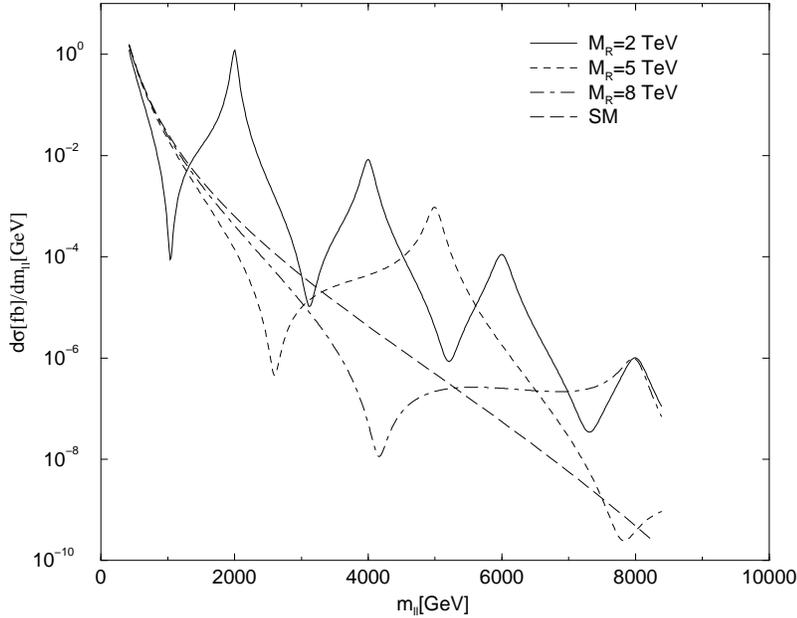}
 \caption{\it
A plot of the differential cross section $d\sigma/dm_{ll}$ as a 
 function of the dilepton invariant mass $m_{ll}$ for the process
$pp\rightarrow l^+l^- +X$ including the effects of Kaluza-Klein 
excitations for the case
when the compactification scale is $M_R=2$ TeV (solid), $M_R=5$ TeV (dashed),  
and  $M_R=8$ TeV (dot-dashed). The resonance structure exhibits the
existence of Kaluza-Klein modes.
For comparison the result of SM case is also
shown (long-dashed).  (Taken from Ref.\cite{nyy}).
\label{exfig1} }
\end{figure}

 The main signal of the Kaluza-Klein  modes is the Drell-Yan process 
  $ pp \rightarrow l^+ l^- +X$ via the 
Kaluza-Klein excitations of the $Z$ and $\gamma$. 
The detailed analysis of the above process yields the  cross section 
for the Kaluza-Klein case which is $\sim 10$ times larger than that 
for the case of the SSM $Z'$ boson. The reason for this 
  enhancement is two fold. First, one has 
the extra factor of $\sqrt{2}$ in the couplings of the 
Kaluza-Klein states to matter as discussed in Sec.4 (see Eq.(1)).
 Second, there is also an
enhancement from a constructive  interference between Kaluza-Klein modes of 
the photon and of the Z  boson which essentially overlap.
There are also other remarkable features associated with the
production of the dilepton pair via Kaluza-Klein states. 
An interesting quantity to plot is the  cross section $d\sigma/dm_{ll}$ 
as a function of the dilepton invariant mass $m_{ll}$ (see Fig.1).
This cross section exhibits  clear resonance 
peaks corresponding to the masses of the Kaluza-Klein  states.
An interesting feature is that Breit-Wigner resonances arising
from the Kaluza-Klein excitation of the photon and from the 
Kaluza-Klein excitation
of the Z boson superpose and lead to a net distorted Breit-Wigner
resonance. Another interesting phenomenon is that there are  
sharp dips below the resonance peaks. 
The origin of these  dips is due to a destructive interference 
between the contributions arising from the exchange of the $\gamma$ and
Z gauge bosons and of their Kaluza-Klein excited states in the region
below the peaks. This phenomenon is unique to the Kaluza-Klein  excitations. 
The analysis shows that Kaluza-Klein excitations with $M_R$ up
to 6 TeV can be explored with a luminosity of 100 
$fb^{-1}$\cite{nyy}.

\begin{figure}[t]
 \vspace{8.0cm}
\includegraphics{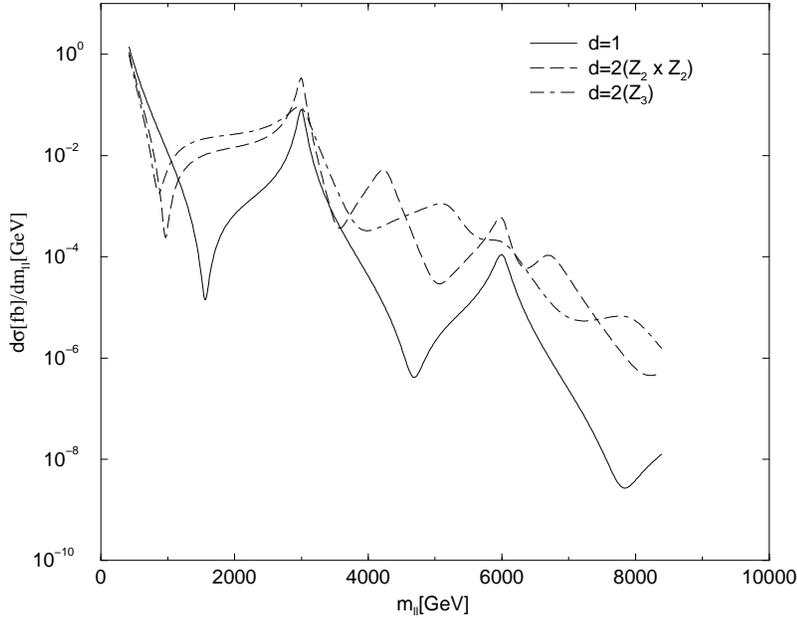}
 \caption{\it
A plot of the differential cross section $d\sigma/dm_{ll}$ as a 
 function of the dilepton invariant mass $m_{ll}$ for the process
$pp\rightarrow l^+l^- +X$ including the effects of
Kaluza-Klein excitations for the case d=1 (solid) and for the
case d=2 for two orbifold compactifications, $Z_2\times Z_2$ 
(dashed) and $Z_3$ (dot-dashed), when the mass of the first 
Kaluza-Klein excitation is taken to be 3 TeV. 
The features of the resonance structure distinguish cases with
different  number of
compactified dimensions as well as cases with  different 
orbifold compactifications.
(Taken from Ref.\cite{nyy}).
\label{exfig2} }
\end{figure}

  Next we consider the case of more than one extra dimension. 
Here for $d>1$ there are in general several orbifold compactifications
 possible and thus the compactifications are more model 
 dependent in this case.
For example, for the case d=2  one can get a 
$Z_2\times Z_2$ orbifold model where the compactified space is 
$S^1/Z_2\times S^1/Z_2$ where the two $S^1$ have a common radius R.
Another possibility is $Z_3$ or $Z_6$ compactification with a 
2D torus of periodicity $2\pi R$. We note in passing that 
the mass spectra of the Kaluza-Klein excitations for the $Z_3$ 
and $Z_6$ orbifold cases are related. Thus for the $Z_3$ orbifold 
compactification masses for the Kaluza-Klein excitations are given
by $M^2_{Z_3}=\frac{4}{3R^2}(m_1^2+m_1m_2+m_2^2)$ where 
$m_1,m_2$ are positive or negative integers. 
Analogously 
for $Z_6$ orbifold compactification the mass fomula for the Kaluza-Klein
excitations is $M^2_{Z_6}=\frac{4}{3R^2}(m_1^2-m_1m_2+m_2^2)$.
We note that the mass formulae for the $Z_3$ and $Z_6$ cases are
related by  $(m_1,m_2)\rightarrow (m_1,-m_2)$.
In general the masses of the  Kaluza-Klein 
excitations, their multiplicities and the strength of their 
couplings to the boundary fermions depend on the nature of the
compactification and these should manifest in the 
production cross-section and in the resonance structure of these states 
at the LHC. A  detailed  analysis of the above bears this out
and one finds that the d=1 and the d=2 compactifications can 
 be distinguished by a detailed 
study of the dileptonic cross section $d\sigma_{ll}/dm_{ll}$ as a 
function of the dilepton invariant mass $m_{ll}$(see Fig.2).
Further as the analysis of Fig.2 shows one can even distinguish
between the $Z_2\times Z_2$ and $Z_3$ compactifications for 
the d=2 case. 
Thus a study of the resonance structure of the
Kaluza-Klein states will allow one to determine the dimensionality of 
the compactified space as well as the detailed nature of 
 specific orbifold compactification. In general the compactification
 radii for different compact dimensions could be different 
 leading to a richer  resonance structure. However, 
 the general observations made above should still hold.  
  A similar analysis can be carried out for the
  study of Kaluza-Klein excitations at future lepton
 colliders which also  present an 
interesting possibility for the study 
of  extra dimensions\cite{rizzo1}.

\section{Low Scale Quantum Gravity}
 As discussed in Sec.3, in addition to the Planck scale being
low, the fundamental scale of gravity may also be low because
the relation between the Planck scale and the fundamental scale
of gravity depends on the number of extra dimensions. 
Thus from  Gauss's law the relation between the volume $R^n$ of 
n new dimensions, the fundamental scale M and the observed Planck 
scale in four dimensions is  
$M_{Pl}^2=R^n M^{n+2}$\cite{a1} where $M_{Pl}=G_N^{-1/2}$.
 One might investigate what happens
if the fundamental scale M of quantum gravity is 1TeV. The case n=1 
is then excluded since it would modify Newtonian law of gravitation
at planetary distances. The case n=2 gives $ R\simeq 1mm$ and
represents an interesting possibility for exploration.
The low energy effective lagrangian for theories of this type is 
discussed in Ref.\cite{sundrum} and 
the implications at accelerators for this class of theories have
been discussed by several authors\cite{peskin}. However,
 astrophysical considerations seem to indicate a limit on the
fundamental scale which is rather large and would seem
to exclude the possibility of observation of quantum gravity
phenomenon at accelerator energies. Thus for the interesting case 
of two extra dimensions
 one finds that the analysis of graviton decay to the cosmic
diffuse gamma radiation\cite{hallsmith} and studies of
 graviton emission into large
compact dimensions from a hot supernova core
using the SN1987A data\cite{cullen} 
 put bounds on the fundamental scale which are very stringent
 , in the range of 50-100 TeV, 
 and place the exploration of extra dimensions  beyond the reach 
of the laboratory  experiment. 
Extra compact dimensions can also
be probed  directly in gravity experiments and
 there are experiments proposed to probe distances at the
 submillimeter scale to look for possible deviations from 
the  inverse square law\cite{long}.
These experiments look for modifications of the type
\begin{equation}
V(r)=-G_N \frac{m_1m_2}{r}(1+\alpha e^{-r/\lambda})
\end{equation}
a form which is valid for $r>>\lambda$ and $\alpha =n+1$ for
n-sphere and $\alpha=2n$ for n-torus compactification\cite{kehagias}. 
For the case of interest of
two extra dimensions $\lambda =R\simeq  1mm$ and $\alpha=3(4)$ 
for the sphere(torus) case. 
The very recent
result from the Seattle Group probes distances well below
the 1mm level and finds no deviation from the inverse square
 law\cite{hoyle}. The experiment places a limit on M of 
$M \geq 3.5$ TeV\cite{hoyle}.

Next we discuss the generation of neutrino masses in models of this
type. In grand unified 
theories and in string theories a small 
neutrino mass is generated by a see-saw mechanism which gives
 $m_{\nu}\sim m_{f}^2/M_X$ where $m_f$ is the fermion mass and $M_X$ is 
a heavy mass scale. In this mechanism the neutrino mass is small because
$M_X$ is heavy, where $M_X$ is taken to lie between the 
intermediate scale and the GUT  scale. 
In a model with large radius compactification 
such a large mass scale does not exist and one needs 
to rethink how a small neutrino mass will arise in such a scenario. 
One mechanism used is to assume that aside from gravity some matter 
fields could
also propagate in the bulk. Specifically it is possible to 
 generate a small Dirac neutrino mass by assuming that the right 
handed component of this neutrino is a Standard Model singlet which 
resides in the bulk\cite{addm,perez}. 
Here the couplings between the singlet and the 
Standard Model particles arise at the wall and the 
Dirac neutrino mass is thus suppressed
because of the volume factor from extra dimensions\cite{addm}. 
However, generation
of a Majorana mass for the neutrino is more difficult as 
 one needs to violate lepton number on a distant 
brane (or in the bulk) and communicate this breaking to the physical 
brane by a bulk field.

A more recent work in a similar direction is on non-factorizable 
geometries\cite{gogber,rs}.
 An example of this  is a 5D model with gravity in the 
bulk and the 5th dimension compactified on $S^1/Z_2$:
$ \{x^M\}=\{x^{\mu},\phi\}$; $\mu=0,..3, -\pi\leq \phi\leq \pi$.
One assumes the existence of two 3-branes one at $\phi=0$
and the other at  $\phi=\pi$. One of these could be viewed as
the brane for the hidden sector and the other as the brane for
the visible sector.
The total action is $S=S_{grav}+S_{vis}+S_{hid}$. 
One looks for solutions with Lorentz invariance with the form
$ds^2=e^{-2\sigma (\phi)}\eta_{\mu\nu}dx^{\mu}dx^{\nu}+r_c^2d\phi^2$.
A fine tuning among the cosmological constants in the bulk and on 
the boundary is necessary, generating a sort of $ADS_5$ geometry,
to achieve a 4D Poincare invariance. Solutions require
  $\sigma (\phi)=k r_c|\phi|$ generating a warp factor
of $e^{-2kr_c|\phi|}$ which decays exponentially as one 
moves away from the wall at $\phi =0$. Some of the phenomenological
consequences of this model are discussed in Refs.\cite{hewett}.

\section{Difficulties in Models with Large Radius Compactifications} 
There are several phenomena other than those discussed above 
 which are affected in scenario
with large radius compactifications creating in some cases extra
 challenges or problems depending on one's point of view.
 We begin with a discussion of the problem of gauge coupling
unification. In MSSM unification of the gauge couplings
takes place naturally with a unification scale of $\sim 2\times 10^{16}$
GeV. In models with large radius compactifications the evolution of
the gauge couplings above the compactification scale obeys
a power law behavior as a function of the 
scale factor\cite{veneziano,d}. This power law behavior arises
as a consequence of the contributions from the Kaluza-Klein excitations.
Thus in models with large radius compactification the meeting of 
 two of the gauge couplings
constants, say $\alpha_1$ and $\alpha_2$,  can occur at a low scale. 
However, with the MSSM spectrum
the low scale unification leads to a value of $\alpha_3$ which 
in general is larger than the one given by the LEP data. Thus one of the
successes of MSSM, i.e., a natural unification of the gauge coupling
constants, is lost. Suggestions on how to recover unification 
with additional contributions are discussed in some of the
papers in Ref.\cite{uni,zk}. However, in this case the unification of 
the gauge couplings becomes more of an accident rather than
 a prediction of the model. 

Perhaps the most serious problem in models with large radius 
compactification is that of proton stability.
Since in theories of large radius compactifications the unification mass
is typically in the TeV range compared to the unification scale of
$\sim 10^{16-17}$ GeV in unified theories of the normal sort
which includes grand unified theories\cite{pgrand} and old 
fashioned string models\cite{pstring}, one has to suppress
baryon and lepton number violating operators to a very high order. 
One suggestion made to overcome this difficulty is to 
assume that the baryon number is gauged in the bulk and 
that this symmetry is then broken on a brane different from the 
physical brane\cite{ad}. In this case one can arrange proton decay 
to be suppressed by a huge exponential factor. 
However, it is not clear how one may naturally arrange the breaking
of baryon number symmetry on the distant brane. Another set of 
suggestions to suppress proton decay require imposition of a
discrete symmetry\cite{d,tye,zk,ellis}. A detailed analysis
of such discrete gauge symmetries is given in Ref.\cite{zk}
where a generalized matter parity of the type $Z_3\times Z_3$ is
proposed in an extended MSSM model which suppresses dangerous 
operators to high orders.
However, it has been argued that unless a theory has an exact or an 
almost exact baryon number conservation one may have rapid proton decay
induced by quantum gravity effects\cite{perry}.
To suppress this type of proton decay one would need a scale 
of quantum gravity which
is similar to the scale one needs to stabilize the proton in 
grand unified theories and in ordinary string unified 
theories\cite{pgrand,pstring}.

\section{Conclusions}
Type I [Type II]  strings allow for the possibility of models
with large radius compactifications.
In this paper we have considered the physical implications of
models where the mass scale $M_R=R^{-1}$ associated with
the extra compactified dimensions is in the TeV region. 
We showed that in this case if
 the accelerator energies are large enough to produce
Kaluza-Klein excitations, then the experimental data 
 can provide information on the number of compactified 
dimensions as well as on the nature of orbifold compactification.
 Such information can be gleaned from a detailed
study of the differential cross section $d\sigma/dm_{ll}$ as a 
function of the dilepton invariant mass $m_{ll}$ from the Drell-Yan process  
$pp\rightarrow l^+l^-+X$. Specifically this process is an important
channel for the discovery of such states up to $M_R\approx 6$ TeV for
an integrated luminosity of 100$fb^{-1}$. Additional  processes   
such as $pp\rightarrow l^{\pm}\nu_l+X$ and $pp\rightarrow jj+X$ also 
provide further  signals for the discovery of Kaluza-Klein modes.
An important unknown in these analyses is the compactification scale 
$M_R$. 
 Currently the strongest constraint on $M_R$  arises from the 
closeness  of the
Standard Model prediction of the Fermi constant $G_F$ and its 
 precision determination from the muon lifetime. 
The Standard Model prediction
depends on the accuracy of the experimental determinations  of 
electro-weak parmeters and with  their current errors one finds 
$M_R\geq 3$ TeV. This limit will increase as the precision of the
electro-weak parameters increases.
Effects of Kaluza-Klein excitations on $a_{\mu}$ are found to be small 
when the constraint from $G_F$ is imposed. Thus if a  deviation 
in $a_{\mu}$ from the Standard Model value is observed in the BNL experiment
it will most likely be an effect other than from 
contributions from the Kaluza-Klein excitations.

 A similar situation holds in the quantum gravity sector
where the recent experiment on the sub-millimeter tests of
 gravity  explores distances well below 
the 1mm level and finds no deviations from the inverse square law.
Interestingly the lower limit on $M$ of $M\geq 3.5$ TeV deduced 
in this experiment is similar to the limit on $M_R$ 
of $M_R\geq 3$ TeV  gotten from the $G_F$ constraint. 
Finally, we note that problems regarding
the consistency of theories with large radius compactifications
persist, the most serious being that of rapid proton decay
in such theories.

\noindent
{\bf Acknowledgements}\\ 
I wish to thank the Physics Institute at the University of Bonn
and the Max Planck Institute, Heidelberg, for hospitality
and acknowledge support from an Alexander von Humboldt award.
The author acknowledges a fruitful collaboration with 
Masahiro Yamaguchi and Youichi Yamada. 
The discussion of Secs.5,6, and 7 is based on work  done in 
collaboration with them. This research was supported in part by 
NSF grant PHY-9901057.

\end{document}